\documentclass[12pt]{article}
\usepackage{amsmath,amssymb,amsthm,bm,cite,fullpage}
\usepackage{graphicx}

\title{Testing one-body density functionals on a solvable model}

\author{C.~L. Benavides-Riveros,$^{1,2}$
and J.~C. V\'arilly$^3$
\\ \\
$^1$Zentrum f\"ur Interdisziplin\"are Forschung, Wellenberg 1
\\
Bielefeld 33615, Germany
\\ \\
$^2$Departamento de F\'isica Te\'orica, Universidad de Zaragoza
\\ 
50009 Zaragoza, Spain
\\ \\
$^3$Escuela de Matem\'atica, Universidad de Costa Rica
\\
San Jos\'e 2060, Costa Rica}

\date{\today}

\newcommand{\dl}{\delta}            
\newcommand{\ga}{\gamma}            
\newcommand{\om}{\omega}            
\renewcommand{\th}{\theta}          
\newcommand{\vs}{\varsigma}         
\newcommand{\vth}{\vartheta}        

\newcommand{\dn}{{\mathord{\downarrow}}} 
\newcommand{\downto}{\downarrow}    
\newcommand{\half}{\tfrac{1}{2}}    
\newcommand{\ketbra}[2]{|#1\rangle\langle#2|} 
\newcommand{\N}{\mathbb{N}}         
\newcommand{\shalf}{{\scriptstyle\frac{1}{2}}} 

\newcommand{\stirling}{\genfrac\{\}{0pt}{1}} 
\newcommand{\up}{{\mathord{\uparrow}}} 
\newcommand{\word}[1]{\quad\mbox{#1}\quad} 
\newcommand{\x}{\times}             

\newcommand{\ee}{\mathrm{ee}}       
\newcommand{\ext}{\mathrm{ext}}     
\newcommand{\kin}{\mathrm{kin}}     
\newcommand{\total}{\mathrm{total}} 

\newcommand{\BBc}{\mathrm{BBC}}     
\newcommand{\BBcuno}{\mathrm{BBC1}} 
\newcommand{\Cuno}{\mathrm{C1}}     
\newcommand{\CGA}{\mathrm{CGA}}     
\newcommand{\CHF}{\mathrm{CHF}}     
\newcommand{\HF}{\mathrm{HF}}       
\newcommand{\GU}{\mathrm{GU}}       
\newcommand{\Mu}{\mathrm{M}}        

\newcommand{\vecform}{\bm}          
\newcommand{\PP}{\vecform{P}}       
\newcommand{\pp}{\vecform{p}}       
\newcommand{\rr}{\vecform{r}}       
\newcommand{\RR}{\vecform{R}}       
\newcommand{\xx}{\vecform{x}}       
\newcommand{\zz}{\vecform{z}}       

\newcommand{\nay}{\textsc{no}}      
\newcommand{\yea}{\textsc{yes}}     

\makeatletter
\def\section{\@startsection{section}{1}{\z@}{-3.5ex plus -1ex minus
 -.2ex}{2.3ex plus .2ex}{\large\bfseries}}
\def\subsection{\@startsection{subsection}{2}{\z@}{-3.25ex plus -1ex
 minus -.2ex}{1.5ex plus .2ex}{\normalsize\bfseries}}
\makeatother

\begin{document}

\maketitle

\begin{abstract}
There are several physically motivated density matrix functionals in
the literature, built from the knowledge of the natural orbitals and
the occupation numbers of the \textit{one-body} reduced density
matrix. With the help of the equivalent phase-space formalism, we
thoroughly test some of the most popular of those functionals on a
completely solvable model.
\end{abstract}

\section{Introduction}
\label{sec:introibo}

At the Colorado conference on Molecular Quantum Mechanics (1959),
Coulson pointed out that in the standard approximation the two-body
density matrix $\ga_2$ carries all necessary information required for
calculating the quantum properties of atoms and molecules
\cite{Coulson60}. Because electrons interact pairwise, the main idea
consists in systematically replacing the quantum wave function by the
two-body reduced matrix (a function of four spatial variables), which
may be obtained by integration of the original $N$-body density matrix
(a function of $2N$ spatial variables). The $N$-representability
problem for the two-body reduced density matrix has proved to be a
major challenge for theoretical quantum chemistry~\cite{Mazziotti12a}.

In the last fifteen years there has been a considerable amount of work
on \textit{Ans\"atze} for the two-body matrix in terms of the one-body
density matrix $\ga_1$. Starting with the pioneer work by M\"uller
\cite{Mueller84}, rediscovered in \cite{BuijseB02}, several competing
functionals have been designed, partly out of theoretical prejudice,
partly with the aim of improving predictions for particular systems:
among others, the total energy of molecular dissociation
\cite{GoedeckerU}, the correlation energy of the homogeneous electron
gas \cite{CsanyiGA02} and the band gap behavior of some 
semiconductors~\cite{SharmaDLG08}.

Two-electron systems are special in the sense that $\ga_2$ can be
reconstructed ``almost exactly'' in terms of~$\ga_1$. Namely, let us
express $\ga_1$ by means of the spectral theorem in terms of its
natural spin orbitals $\{\phi_i(\rr)\}$ and its occupation numbers
$\{n_i\}$. The ground state of this system (which is of closed-shell
type) admits a one-density matrix:
\begin{equation}
\ga_1(\xx,\xx') = \bigl( \up_1\up_{1'} + \dn_1\dn_{1'} \bigr)
\ga_1(\rr,\rr') = \bigl( \up_1\up_{1'} + \dn_1\dn_{1'} \bigr)
\sum_i n_i \, \phi_i(\rr)\phi_i^*(\rr'),
\label{eq:one-DM} 
\end{equation}
where $\xx = (\rr,s)$ stands for the spatial and spin coordinates. The
natural occupation numbers satisfy $\sum_i n_i = 1$, with
$0 \leq n_i \leq 1$. Mathematically this is a mixed state. The
original two-density matrix is given by the
Shull--L\"owdin--Kutzelnigg (SLK)
formula~\cite{LoewdinS56,Kutzelnigg63}:
\begin{gather}
\ga_2(\xx_1,\xx_2;\xx'_1,\xx'_2) 
= \bigl( \up_1\dn_2 - \dn_1\up_2 \bigr)
\bigl( \up_{1'}\dn_{2'} - \dn_{1'}\up_{2'})
\sum_{ij} \frac{c_ic_j}{2}\, \phi_i(\rr_1) \phi_i(\rr_2) 
\phi_j^*(\rr'_1) \phi_j^*(\rr'_2),
\notag \\
\word{with coefficients} c_i = \pm \sqrt{n_i} \,.
\label{eq:two-DM} 
\end{gather}
Though the expression is exact, the signs of the $c_i$ still need to
be determined to find the ground state. Note that
\begin{equation}
\ga_1(\xx;\xx') 
= 2 \int \ga_2(\xx,\xx_2;\xx',\xx_2) \, d\xx_2 .
\label{eq:sum-rule} 
\end{equation}
This condition becomes a \textit{sum rule} which, as we shall see, may
be satisfied or not by proposed density matrix functionals. Among
other conditions, $\ga_2$ is Hermitian: 
$\ga_2(\xx_1,\xx_2;\xx'_1,\xx'_2) =
\ga_2^*(\xx'_1,\xx'_2;\xx_1,\xx_2)$, and antisymmetric in each pair of
subindices:
$$
\ga_2(\xx_1,\xx_2;\xx'_1,\xx'_2) = -\ga_2(\xx_2,\xx_1;\xx'_1,\xx'_2) 
= -\ga_2(\xx_1,\xx_2;\xx'_2,\xx'_1).
$$

In the early years of the theory, Heisenberg invented an exactly
solvable model, here called \textit{harmonium}, as a proxy for the
spectral problem of two-electron atoms~\cite{Heisenberg26}. It
exhibits two fermions interacting with an external harmonic potential
and repelling each other by a Hooke-type force; its Hamiltonian, in
Hartree-like units, is
\begin{equation}
H = \frac{|\pp_1|^2}{2} + \frac{|\pp_2|^2}{2}
+ \frac{k}{2}(|\rr_1|^2 + |\rr_2|^2) - \frac{\dl}{4} r_{12}^2,
\label{eq:Mosh-atom} 
\end{equation}
where $\dl>0$ and $r_{12} := |\rr_1 - \rr_2|$. Many years later,
Moshinsky~\cite{Moshinsky68} came back to it with the purpose of
calibrating correlation energy ---see also
\cite{MarchCCA08,Loos10,NagyP11}. Also, Srednicki~\cite{Srednicki93}
used the harmonium model to study the black hole entropy, proving its
proportionality to the black hole area.

Recently, within the context of a phase-space density functional
theory~\cite{Dahl09}, here called WDFT, the alternating choice of
signs in~\eqref{eq:two-DM} has been shown to be the correct one for
the harmonium ground state~\cite{Pluto,Pallene,Hermione}.
Mathematically, density matrix functional theory (DMFT) and density
functional theory on phase space are equivalent: see the next section.
Thus, its density matrix functional \eqref{eq:two-DM} is nowadays
known exactly. Some of lower excited configurations of harmonium are
also of much current interest
\cite{YanyezPD10,BouvrieMPSMD12,Laetitia}.

The integrability and solvability of harmonium enables one to test
accurately how proposed density matrix functionals behave for this
particular system~\cite{AmovilliM03}. There is another particularity
of harmonium: for its ground state, the M\"uller functional, evaluated
on the exact one-body reduced density matrix, yields the correct value
of the energy~\cite{NagyP10}. Here we confirm by a different method
this surprising coincidence, and we catalogue the predictions for
harmonium by several proposed two-body functionals, measured against
the exact model.

\vspace{6pt}

In section~\ref{sec:harmonium} we briefly recall the analytical
phase-space treatment for the harmonium ground state; this also helps
to introduce the notation. In section \ref{sec:HFM} the Hartree--Fock
and M\"uller functionals of the one-body density are discussed in the
context of harmonium. Section~\ref{sec:testing} starts the systematic
comparison of several other approximate functionals proposed in the
literature; we compute the error in the interelectronic energy value
given by each of them for the \textit{exact} family of ground states
parametrized by $(k,\dl)$. They all behave worse than M\"uller's. We
append some concluding remarks.

\section{Wigner natural orbitals for the harmonium ground state}
\label{sec:harmonium}

The basic object of WDFT for a two-electron atom is the one-body
quasiprobability $d_1$. Let us look at the two-body quasiprobability
$d_2$. Given any interference operator $\ketbra{\Psi}{\Phi}$ acting on
the Hilbert space of the two-electron system, we denote
\begin{align}
& P_{2\,\Psi\Phi}(\rr_1,\rr_2; \pp_1,\pp_2;
\vs_1,\vs_2;\vs_{1'},\vs_{2'})
\label{eq:Wigner-trans} 
\\
&\quad := \int \Psi(\rr_1-\zz_1, \rr_2-\zz_2;\vs_1,\vs_2) \,
\Phi^*(\rr_1+\zz_1, \rr_2+\zz_2;\vs_{1'},\vs_{2'}) \,
e^{2i(\pp_1\cdot\zz_1 + \pp_2\cdot\zz_2)} \, d\zz_1\,d\zz_2.
\notag
\end{align}
These are $4 \x 4$ matrices on spin space. When the interference
operator corresponds to a pure state ($\Psi = \Phi$) we speak of
\textit{Wigner quasiprobabilities}. In this case, the functions
\eqref{eq:Wigner-trans} are real, and we write $d_2$ for~$P_2$. Its
integral equals~$1$. The extension of this definition to mixed states
is immediate. The corresponding reduced one-body functions are found
by integration:
\[
P_{1\,\Psi\Phi}(\rr_1;\pp_1;\vs_1;\vs_{1'}) = 2 \int
P_{2\,\Psi\Phi}(\rr_1,\rr_2;\pp_1,\pp_2; \vs_1,\vs_2;\vs_{1'},\vs_2)
\,d\rr_2 \,d\pp_2 \,d\vs_2.
\]
On spin space these are $2\x2$ matrices. When $\Psi = \Phi$ we write
$d_1$ for~$P_1$. The integral of this quantity equals~$2$. The
associated spinless quantities are obtained by tracing on the spin
variables:
\begin{align*}
d_2(\rr_1,\rr_2; \pp_1,\pp_2) &= \int d_2(\rr_1,\rr_2; \pp_1,\pp_2;
\vs_1,\vs_2;\vs_1,\vs_2) \,d\vs_1 \,d\vs_2
\\
\word{and} d_1(\rr;\pp) 
&= \int d_2(\rr,\rr_2; \pp,\pp_2) \,d\rr_2 \,d\pp_2 .
\end{align*}
The marginals of $d_2$ give the pairs densities $\rho_2(\rr_1,\rr_2)$,
$\pi_2(\pp_1,\pp_2)$. The marginals of $d_1$ give the electronic
density, namely $\rho(\rr) = \int d_1(\rr,\pp) \,d\pp$, and the
momentum density $\pi(\pp) = \int d_1(\rr,\pp) \,d\rr$. It should be
obvious how to extend the definitions to $N$-electron systems and
their reduced quantities; the combinatorial factor for
$d_N \mapsto d_n$ is~$\binom{N}{n}$.

Putting together the equations \eqref{eq:two-DM} and \eqref{eq:one-DM}
with~\eqref{eq:Wigner-trans}, one arrives~\cite{Pluto} at:
\begin{align}
d_2(\rr_1,\rr_2;\pp_1,\pp_2;\vs_1,\vs_2;\vs_{1'},\vs_{2'})
&= (\text{spin factor}) \x \sum_{ij} \frac{c_i\,c_j}2 \,
\chi_{ij}(\rr_1;\pp_1) \chi_{ij}(\rr_2;\pp_2),
\label{eq:two-density} 
\\[-\jot]
\text{and}\quad  d_1(\rr_1;\pp_1;\vs_1,\vs_{1'})
&= 2 \int d_2(\rr_1,\rr_2;\pp_1,\pp_2; \vs_1,\vs_2;\vs_{1'},\vs_2)
\,d\vs_2 \,d\rr_2 \,d\pp_2
\notag \\
&= \bigl( \up_1\up_{1'} + \dn_1\dn_{1'} \bigr)
\sum_i n_i\, \chi_i(\rr_1;\pp_1),
\notag
\end{align}
where $n_i$ are the occupation numbers with $0 \leq n_i \leq 1$ and
$\sum_i n_i = 1$, the $\chi_{ij}$ the natural Wigner interferences and
$\chi_i := \chi_{ii}$ denote the natural Wigner orbitals; the spin
factor is that of~\eqref{eq:two-DM}. Evidently
$\bigl( \up_1\up_{1'} + \dn_1\dn_{1'} \bigr)$ is a rotational scalar;
thus we replace it by~$2$ in what follows.

Introducing extracule and intracule coordinates, respectively given by
\begin{align*}
\RR &= \frac{1}{\sqrt{2}}(\rr_1 + \rr_2),  \qquad
\rr = \frac{1}{\sqrt{2}}(\rr_1 - \rr_2),
\\
\PP &= \frac{1}{\sqrt{2}}(\pp_1 + \pp_2),  \qquad
\pp = \frac{1}{\sqrt{2}}(\pp_1 - \pp_2),
\end{align*}
the harmonium Hamiltonian \eqref{eq:Mosh-atom} is rewritten:
\[
H = H_R + H_r
:= \frac{P^2}{2} + \frac{\om^2 R^2}{2} + \frac{p^2}{2}
+ \frac{\mu^2 r^2}{2},
\]
where $\om := \sqrt k$ and $\mu := \sqrt{k - \dl}$. We assume
$\dl < k$. The energy spectrum for harmonium is obviously
$(\N + \frac{3}{2})\om + (\N + \frac{3}{2})\mu$ and the energy of the
ground state is $E_0 = \frac{3}{2}(\om + \mu)$. For this 
configuration, the (spinless) Wigner two-body quasiprobability is
readily found~\cite{Dahl09}:
\begin{align}
d_2(\rr_1,\rr_2;\pp_1,\pp_2)
&= \frac{1}{\pi^6} \exp\biggl(-\frac{2H_R}{\om} \biggr) 
\exp\biggl( -\frac{2H_r}{\mu} \biggr).
\label{eq:harmon-d2} 
\end{align}
The reduced one-body phase-space quasiprobability for the ground state
is thus obtained:
\[
d_1(\rr;\pp) = 2 \int d_2(\rr,\rr_2;\pp,\pp_2) \,d\rr_2 \,d\pp_2
= \frac{2}{\pi^3} \biggl( \frac{4\om\mu}{(\om + \mu)^2} \biggr)^{3/2}
e^{-2r^2\om\mu/(\om + \mu)} e^{-2p^2/(\om + \mu)}.
\]
Its natural orbital expansion, with $i$ integer $\geq 0$ and $L_i$
the corresponding Laguerre polynomial, reads~\cite{Pluto}:
\begin{align}
c_i^2 = n_i 
&= \frac{4\sqrt{\om\mu}}{\bigl(\sqrt\om + \sqrt\mu\,\bigr)^2}
\biggl(\frac{\sqrt\om - \sqrt\mu}{\sqrt\om + \sqrt\mu} \biggr)^{2i}
=: (1 - t^2)\,t^{2i} \,;
\label{eq:harmon-param} 
\\
\chi_i(\rr_1;\pp_1) &= \chi_i(x_1;p_{1x}) \chi_i(y_1;p_{1y})
\chi_i(z_1;p_{1z}), \word{where}
\notag \\
\chi_i(x;p_x) &= \frac{1}{\pi}\, (-1)^i
L_i\bigl( 2\sqrt{\om\mu}\,x^2 + 2p_x^2/\sqrt{\om\mu} \bigr) 
e^{-\sqrt{\om\mu}\,x^2 - p_x^2/\sqrt{\om\mu}}.
\notag
\end{align}
Up to a phase, the functions $\chi_i$ determine the set of
interferences: for $j \geq k$,
\begin{align}
\chi_{jk}(x,p_x) 
&= \frac{1}{\pi}\, (-1)^k \frac{\sqrt{k!}}{\sqrt{j!}}\,
\bigl(2\sqrt{\om\mu}\,x^2 + 2p_x^2/\sqrt{\om\mu} \bigr)^{(j-k)/2}
\nonumber \\
&\qquad \x e^{-i(j - k)\vth}
L_k^{j-k} \bigl( 2\sqrt{\om\mu}\,x^2 + 2p_x^2/\sqrt{\om\mu} \bigr)
e^{-\sqrt{\om\mu}\,x^2 - p_x^2/\sqrt{\om\mu}},
\label{eq:Lag-basis} 
\end{align}
where $\vth := \arctan(p_x/\!\sqrt{\om\mu}\,x)$. The $L_k^{j-k}$ are
associated Laguerre polynomials. The $\chi_{kj}$ are complex
conjugates of the~$\chi_{jk}$.

\vspace{6pt}

The SLK relation $c_i = \pm \sqrt{n_i}$ must hold. There is the
problem of determining the signs of this infinite set of square roots,
so as to find the ground state. In principle, to recover $d_2$ from
$d_1$ is no mean feat, since it involves going from a statistical
mixture to a pure state. The advantage in the present case is that the
result \eqref{eq:harmon-d2} is simple, particularly so on phase space,
and known. With the \textit{alternating choice} (unique up to a global
sign):
\begin{equation}
c_i = (-)^i\,\sqrt{n_i} = \sqrt{1 - t^2}\,(-t)^i,
\label{eq:alter-signs} 
\end{equation}
and the above $f_{jk}$, formula \eqref{eq:two-density} does reproduce
\eqref{eq:harmon-d2}. This was originally proved in~\cite{Pluto} by
organizing the series \eqref{eq:two-density} in a square array and
summing over subdiagonals; some special function identities come in
handy at the end.

Incidentally, new special function identities certainly lurk here: a
natural idea in this context is to try to sum the SLK series
differently. Consider for instance the sum on the first column:
\begin{align*}
S_0 := \sum_r \frac{c_rc_0}{2} \,
\chi_{r0}(\rr_1;\pp_1) \chi_{r0}(\rr_2;\pp_2)
&= \frac{1 - t^2}{\pi^2}\, e^{-(U_1^2 + U_2^2)}
\sum_{r=0}^\infty (-1)^r t^r \frac{1}{r!} (4U_1^2U_2^2)^{r/2}
e^{-ir(\vth_1 + \vth_2)}
\\[\jot]
&= \frac{1 - t^2}{\pi^2}\, e^{-(U_1^2 + U_2^2)}
\sum_{r=0}^\infty \frac{1}{r!}
\bigl( -2tU_1U_2 e^{-i(\vth_1 + \vth_2)} \bigr)^r
\\
&= \frac{1-t^2}{\pi^2}\, e^{-(U_1^2 + U_2^2)}
\exp \bigl( -2tU_1U_2 e^{-i(\vth_1 + \vth_2)} \bigr),
\end{align*}
with the notations
$$
U_i := \bigl[\sqrt{\om\mu}\,r_i^2 + p_i^2/\sqrt{\om\mu}\bigr]^{1/2},
\quad  \vth_i := \arctan(p_i/\!\sqrt{\om\mu}\,r_i),  \qquad  i = 1,2.
$$
As recalled in~\cite{SolomonDBHP11}, the exponential generating
function of the Bell polynomials $B_n(y)$ is
$\exp(y(e^x - 1)) = \sum_{n=0}^\infty B_n(y)\,x^n/n!$. The polynomials
are given by
$B_n(y) = \sum_{m=0}^n \stirling{n}{m}\,y^m$, where $\stirling{n}{m}$
is the number of partitions of $n$ into exactly $m$ subsets. Therefore,
$$
\exp(y e^x) = e^y \sum_{n=0}^\infty B_n(y)\, \frac{x^n}{n!} \,;
$$
and so
$$
S_0 = \frac{1 - t^2}{\pi^2}\, e^{-(U_1^2 + 2tU_1U_2 + U_2^2)}
\sum_{n=0}^\infty B_n(-2tU_1U_2) \frac{(-i(\vth_1 + \vth_2))^n}{n!}\,.
$$
We have not found a clear way ahead, however, for the summation of all
columns. On the other hand, the true and tested minimization method
works fine to derive \eqref{eq:alter-signs}, too~\cite{Pallene}.
Trivially, the same sign rule holds for natural orbitals of the garden
variety~\eqref{eq:two-DM}. This is invoked in~\cite{NagyA11} without
explanation.

\section{Hartree--Fock and M\"uller functionals of the one-body density}
\label{sec:HFM}

The total energy of an electronic system is of the form
$$
E_\total[d_1,d_2] = E_\kin[d_1] + E_\ext[d_1] + E_\ee[d_2],
$$
where the kinetic $E_\kin$ and potential $E_\ext$ energies are known
functionals of the spinless one-body Wigner reduced quasiprobability.
In our case, recalling \eqref{eq:harmon-param}:
\begin{align*}
E_\kin[d_1] &= \int \frac{p^2}{2}\, d_1(\rr;\pp)\,d\rr\,d\pp
= \frac{3\om}{4} + \frac{3\mu}{4} = \frac{3\om}{4}
\biggl[ 1 + \Bigl( \frac{1 - t}{1 + t} \Bigr)^2 \biggr];
\\
E_\ext[d_1] &= \int \frac{\om^2r^2}{2}\, d_1(\rr;\pp) \,d\rr\,d\pp
= \frac{3\om}{4} + \frac{3\om^2}{4\mu} = \frac{3\om}{4}
\biggl[ 1 + \Bigl( \frac{1 + t}{1 - t} \Bigr)^2 \biggr].
\end{align*}
The interelectronic repulsion energy $E_\ee[d_2]$ is a functional of
the two-body Wigner quasiprobability or indeed only of the pairs
density:
\begin{align*}
E_\ee[d_2]
&= - \frac{\dl}{4} \int d_2(\rr_1,\rr_2;\pp_1,\pp_2) r_{12}^2
\,d\rr_1\,d\rr_2 \,d\pp_1\,d\pp_2 
\\
= E_\ee[\rho_2]
&= - \frac{\dl}{4} \int\rho_2(\rr_1,\rr_2)r_{12}^2 \,d\rr_1\,d\rr_2 
= \frac{3\mu}{4} - \frac{3\om^2}{4\mu}
= - \frac{3\om}{4}\,\frac{8t(1 + t^2)}{(1 - t^2)^2} \,.
\end{align*}
We note that the kinetic energy of the system stays finite from 
$\mu = \om$ ($t = 0$) to $\mu = 0$ ($t = 1$). The potential energies
diverge as $\mu \downto 0$, in the strong repulsion regime; but their
sum remains finite and equal to the kinetic energy, as prescribed by
the virial theorem.

In the language of this paper DMFT amounts to the search for
functionals for $\rho_2$ in terms of~$d_1$, expressed through its
Wigner natural orbitals and their occupation numbers. We have already
indicated that the wide variety of functionals currently used in DMFT
for computational purposes can be traced back to the functional
proposed by M\"uller~\cite{Mueller84}. Note first that, with an
obvious notation, the \textit{exact} phase-space functional for the
present system is:
\begin{align}
\rho_2(\rr_1,\rr_2) &= \sum_{i,j\geq 0} (-)^{i+j} \sqrt{n_in_j}\,
\chi_{ij}(\rr_1) \chi_{ij}(\rr_2)
\label{eq:exact-r2} 
\\
&= \sum_{i\geq 0} n_i\, \rho_i(\rr_1) \rho_i(\rr_2)
+ \sum_{i\neq j\geq 0} (-)^{i+j} \sqrt{n_in_j}\,\chi_{ij}(\rr_1)
\chi_{ij}(\rr_2),
\notag
\end{align}
with $\rho_i(\rr) \equiv \chi_i(\rr)$ being the electronic density for
the natural orbital $\chi_i(\rr,\pp)$. This is correctly normalized by
$\int \rho_2(\rr_1,\rr_2) \,d\rr_1\,d\rr_2 = \sum_i n_i = 1$, in view
of $\int \chi_{ij}(\rr) \,d\rr = 0$ when $i \neq j$. Translated into
our language, the M\"uller functional for the singlet is of the form
\begin{align}
\rho_2^\Mu(\rr_1,\rr_2)
&= 2 \sum_i n_i\,\chi_i(\rr_1) \sum_j n_j\,\chi_j(\rr_2) 
- \sum_{i,j} \sqrt{n_i\,n_j}\, \chi_{ij}(\rr_1) \,\chi_{ji}(\rr_2)
\label{eq:Mueller-r2} 
\\
&= \frac{1}{2}\, \rho(\rr_1) \rho(\rr_2)
- \sum_i n_i\,\rho_i(\rr_1) \rho_i(\rr_2)
- \sum_{i\neq j} \sqrt{n_i\,n_j}\, \chi_{ij}(\rr_1)\,\chi_{ij}(\rr_2).
\notag
\end{align}
We have used that $\chi_{ij}(\rr) = \chi_{ji}(\rr)$ for real orbitals.
More generally, M\"uller considered $n_i^p\,n_j^q$ with $p + q = 1$
instead of $\sqrt{n_i\,n_j}$. Recently, the case
$\half \leq p = q \leq 1$ has been studied~\cite{SharmaDLG08}. The
M\"uller functional satisfies some nice properties; among them, the
sum rule~\eqref{eq:sum-rule} and hermiticity. For Coulombian systems
its energy functional is convex~\cite{FrankLSS07}. Nonetheless,
antisymmetry fails. (We summarize properties fulfilled or infringed by
each functional in Table~\ref{table:props}.)

\vspace{6pt}

Following Lieb~\cite{Lieb81}, the Hartree--Fock approximation may be
regarded as yet another functional of~$d_1$. This is given by
\begin{align}
\rho_2^\HF(\rr_1,\rr_2) = \frac{1}{2}\, \rho(\rr_1)\rho(\rr_2)
- \sum_i n^2_i \,\rho_i(\rr_1) \rho_i(\rr_2)
- \sum_{i\neq j} n_i n_j \,\chi_{ij}(\rr_1) \,\chi_{ij}(\rr_2).
\label{eq:HF-r2} 
\end{align}
Expressions~\eqref{eq:exact-r2} and~\eqref{eq:HF-r2} coincide only
when the occupation numbers are pinned to $0$ or~$1$. The cumulant
$\rho_2 - \rho^\HF_2$ can be also computed~\cite{Piris06}. The
``best'' Hartree--Fock state, in the sense of best approximation for
the ground state energy with only one $n_i \neq 0$, is
given~\cite{Hermione} by:
\begin{align*}
P_\HF(\rr_1,\rr_2;\pp_1,\pp_2)
&= \frac{1}{\pi^6}\, e^{-(r_1^2 + r_2^2)\sqrt{(\om^2 + \mu^2)/2}}
\, e^{-(p_1^2 + p_2^2)/\sqrt{(\om^2 + \mu^2)/2}}
\\
&= \frac{1}{\pi^6}\, e^{-(R^2 + r^2)\sqrt{(\om^2 + \mu^2)/2}}
e^{-(P^2 + p^2)/\sqrt{(\om^2 + \mu^2)/2}} \,.
\end{align*}
Use of the energy formulas for this state yields
$$
E_\HF = 3\sqrt{(\om^2 + \mu^2)/2},
$$
and so the \textit{correlation} energy is
\begin{align*}
E_c(\om,\mu) &:= E_0 - E_\HF
= \frac{3}{2} \bigl( \om + \mu - \sqrt{2(\om^2 + \mu^2)}\, \bigr)
\\
\word{so that} E_c(\om,0) &= - \frac{3}{2} (\sqrt{2} - 1)\om.
\end{align*}
For small values of $\mu$, however, minimization by use
of~\eqref{eq:HF-r2} gives lower values of the energy than $E_\HF$
\cite{NagyP11}: the results by Lieb on the Hartree--Fock
\textit{functional} for arbitrary states of Coulombian systems do not
apply here. However, it should be remembered that the sum rule fails
for non-Hartree--Fock states.

\section{Exact vs.\ approximate functionals for harmonium ground state}
\label{sec:testing}

Our strategy henceforth is simply to gauge the worth of the
functionals by computing their respective values on the \textit{true}
ground state. As mentioned above, it has recently been found
\cite{NagyP10} that the M\"uller functional yields \textit{precisely}
the correct energy values for harmonium when evaluated on the exact
ground state ---thus, for $N = 2$, it is also \textit{overbinding} for
the harmonic repulsion just as for the Coulombian one
\cite{FrankLSS07,Pluto}, since the minimizing state for that
functional will yield a lower value of the energy.

Thus a feasible procedure is to compute the difference between the
values given by the M\"uller functional and each of the several
functionals whose accuracy we want to study. We need only worry about
the interelectronic repulsion energy; since all the relevant
quantities factorize, for notational simplicity we shall work in
dimension one.

\subsection{M\"uller interelectronic energy} 
\label{ssc:Muller-fnal}

{}From \eqref{eq:Mueller-r2} and~\eqref{eq:Lag-basis}, we get:
\begin{align*}
& \sum_{r,s=0}^\infty \sqrt{n_r n_s}\,
\chi_{rs}(r_1,p_1) \chi_{sr}(r_2,p_2)
\\
&\qquad = \frac{1 - t^2}{\pi^2}\, e^{-(U_1^2 + U_2^2)} \sum_{l=r-s}
(2 U_1 U_2 t)^l \,e^{-il\th} \sum_{s\geq 0} \frac{s!}{(l+s)!}\,
L_s^l(2U_1^2)\, L_s^l(2U_2^2)\, t^{2s}
\\ 
&\qquad = \frac{1}{\pi^2}\, e^{-(U_1^2 + U_2^2)(1 + t^2)/(1 - t^2)} 
\sum_{l=-\infty}^\infty e^{-il\th}
I_l\biggl( \frac{4U_1U_2\,t}{1 - t^2} \biggr)
\\
&\qquad = \frac{1}{\pi^2}\, e^{-(U_1^2 + U_2^2)(1 + t^2)/(1 - t^2)}\,
e^{4U_1U_2\,t \cos\th/(1 - t^2)}
\end{align*}
where $\th = \vth_1 - \vth_2$; some well-known properties of Laguerre
polynomials and modified Bessel functions have been invoked. In all,
the spinless phase-space M\"uller functional for the harmonium ground 
state is, using the notation $u_i := (r_i,p_i)$, 
\begin{align}
& d_2^\Mu(u_1,u_2) 
\label{eq:harmon-d2M} 
\\
&\quad = \frac{2}{\pi^2} \biggl( \frac{1 - t^2}{1 + t^2} \biggr)^2
e^{-(U_1^2 + U_2^2)(1 - t^2)/(1 + t^2)} 
- \frac{1}{\pi^2}\, e^{-(U_1^2 + U_2^2)(1 + t^2)/(1 - t^2)}\,
e^{4U_1U_2\,t \cos\th/(1 - t^2)}.
\nonumber 
\end{align}

\begin{figure}[t] 
\centering
\begin{minipage}[t]{.49\textwidth}
\begin{center}
\includegraphics[width=8cm]{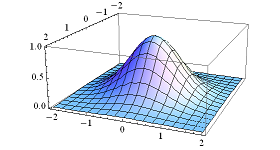} 
\end{center}
\end{minipage}
\hfill
\begin{minipage}[t]{.49\textwidth}
\begin{center}
\includegraphics[width=8cm]{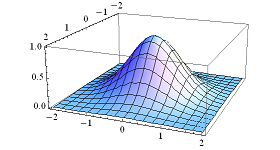}  
\end{center}
\end{minipage}
\\
\begin{minipage}[t]{.49\textwidth}
\begin{center}
\includegraphics[width=8cm]{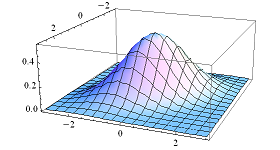} 
\end{center}
\end{minipage}
\hfill
\begin{minipage}[t]{.49\textwidth}
\begin{center}
\includegraphics[width=8cm]{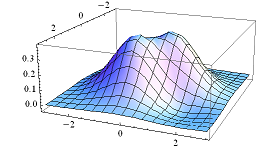}  
\end{center}
\end{minipage}
\\
\begin{minipage}[t]{.49\textwidth}
\begin{center}
\includegraphics[width=8cm]{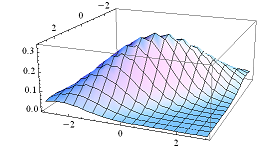} 
\end{center}
\end{minipage}
\hfill
\begin{minipage}[t]{.49\textwidth}
\begin{center}
\includegraphics[width=8cm]{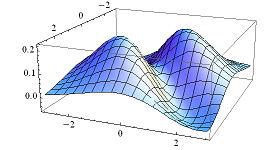}  
\end{center}
\end{minipage}
\hfill
\caption{Exact vs.\ M\"uller pairs density for harmonium at
$t = 0$, $t = 0.3$ and $t = 0.5$. The graphics show the dimensionless
functions $\pi\rho_2/\om$ (on the left) and $\pi\rho_2^\Mu/\om$ (on 
the right) in terms of $\om^{1/2}\,r_1$ and $\om^{1/2}\,r_2$.}
\label{graf:MvsE} 
\end{figure}

In order to compute the interelectronic energy, we proceed with the
mean value of the electronic separation:
$\int r_{12}^2 \,d_2^\Mu(u_1,u_2) \,du_1\,du_2$. For the first term
in~\eqref{eq:harmon-d2M}, we get:
\begin{align*}
\frac{2}{\pi^2}\, \biggl( \frac{1 - t^2}{1 + t^2} \biggr)^2
& \int r_{12}^2 \, e^{-(U_1^2 + U_2^2)(1 - t^2)/(1 + t^2)}
\,du_1 \,du_2
\\ 
&= \frac{1}{\pi}\, \frac{4\om\mu}{\om + \mu} \int r_{12}^2 \,
e^{-2\om\mu(r_1^2 + r_2^2)/(\om + \mu)} \,dr_1 \,dr_2 
= \frac{\om + \mu}{\om\mu} \,.
\end{align*}
For the second term, we obtain:
\begin{align*}
& - \frac{1}{\pi^2} \int e^{-(U_1^2 + U_2^2)(1 + t^2)/(1 - t^2)} \,
e^{4U_1U_2\,t\cos\th/(1 - t^2)} \,dp_2 \,dp_1
\\
&\qquad = - \frac{\sqrt{\om\mu}}{\pi} \int r_{12}^2 \,
e^{-\shalf(r_1^2 + r_2^2)(\om + \mu)} \, e^{r_1r_2(\om - \mu)}
\,dr_2 \,dr_1 = - \frac{1}{\om} \,.
\end{align*}
In the process we have obtained a sort of (spinless) ``M\"uller pairs
density'' for the true ground state,
\begin{equation}
\rho_2^\Mu(r_1,r_2) := \frac{1}{\pi}\, \frac{4\om\mu}{\om + \mu}\,
e^{-2\om\mu(r_1^2 + r_2^2)/(\om + \mu)}
- \frac{\sqrt{\om\mu}}{\pi}\, e^{-\shalf(\om + \mu)(r_1^2 + r_2^2)} \,
e^{(\om - \mu)r_1r_2} \,;
\label{eq:harmon-r2M} 
\end{equation}
whose predicted mean square value for the distance between the two
electrons is
$$
\int r_{12}^2 \,\rho_2^\Mu(u_1,u_2) \,du_1 \,du_2
= \frac{\om + \mu}{\om\mu} - \frac{1}{\om} = \frac{1}{\mu} \,.
$$
The \textit{same} mean square value is easily obtained from the
\textit{exact} pairs density~\cite{Pluto}:
\begin{equation}
\rho_2(r_1,r_2) = \frac{\sqrt{\om\mu}}{\pi}\, 
 e^{-\shalf(\om + \mu)(r_1^2 + r_2^2)} \, e^{(\mu - \om)r_1r_2}.
\label{eq:harmon-exact-rtwo} 
\end{equation}
Thus, both energies \textit{coincide}:
$E_\ee = E_\ee^\Mu = -\dl/4\mu = (\mu^2 - \om^2)/4\mu$. Note that the
result is valid for any value of~$t$. This is surprising because the
shapes of $\rho_2$ and $\rho_2^\Mu$ grow very distinct as $t$
increases ---see Figure~\ref{graf:MvsE}.

\begin{figure}[t] 
\centering
\begin{minipage}[t]{.49\textwidth}
\begin{center}
\includegraphics[width=8cm]{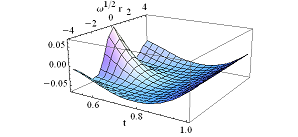} 
\end{center}
\end{minipage}
\hfill
\begin{minipage}[t]{.49\textwidth}
\begin{center}
\includegraphics[width=8cm]{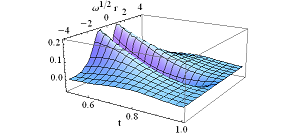}  
\end{center}
\end{minipage}
\hfill \caption{Diagonal part $\pi\rho_2^\Mu(r,r)/\om$ and
antidiagonal part $\pi\rho_2^\Mu(r,-r)/\om$ of the M\"uller functional
as functions of $t$ and $\om^{1/2}\,r$.}
\label{graf:diagonals} 
\end{figure}

In summary, by a somewhat different method, we have confirmed the
result of~\cite{NagyP10}. The coincidence does not hold for other
values $p,q \neq \half$ in the M\"uller approach. It may be considered
fortuitous, because \eqref{eq:harmon-r2M} and the exact pairs density
\eqref{eq:harmon-exact-rtwo} are rather dissimilar: for $t > 0$, the
spinless two-body M\"uller functional does not have a maximum at the
origin in phase space, whereas the exact functional does. More
precisely, as figures \ref{graf:MvsE} and \ref{graf:diagonals} show,
the M\"uller functional exhibits two maxima located at the
antidiagonal sector of the density. Also, it sports negative values at
some points. As pointed out in the original paper~\cite{Mueller84},
this phenomenon is a consequence of the inequality
$\sqrt{n_j} \geq n_j$ satisfied by the natural occupation numbers of
the system. Figures \ref{graf:MvsE} and \ref{graf:diagonals} display
the negativity around the diagonal elements of the density. This
indicates that the M\"uller functional is also unphysical, in a
subtler way than the Hartree--Fock functional~\cite{Helbig06}.

\subsection{Hartree--Fock interelectronic energy} 
\label{ssc:HF-fnal}

We use the following terms, computed in~\cite{Hermione}:
\begin{align*}
L_i &:= \frac{\mu^2 - \om^2}{4} \int \chi_i(r_1)\, (r_1 - r_2)^2 \,
\chi_i(r_2) \,dr_1 \, dr_2 
= (2i + 1)\, \frac{\mu^2 - \om^2}{4\sqrt{\om\mu}}\,,
\\
M_i &:= \frac{\mu^2 - \om^2}{4} \int
\chi_{i,i+1}(r_1) \, \chi_{i+1,i}(r_2)(r_1 - r_2)^2 \,dr_1 \,dr_2
= -(i + 1) \frac{\mu^2 - \om^2}{4\sqrt{\om\mu}} \,.
\end{align*}
The difference between the interelectronic energy predicted by the
Hartree--Fock functional \eqref{eq:HF-r2} and that predicted by the
M\"uller functional on the true harmonium ground states is then given
by:
\begin{align*}
E_\ee^\HF(t) - E_\ee^\Mu(t) 
&= - \frac{\mu^2 - \om^2}{4} \sum_{i,j=0}^\infty
\bigl[ n_i n_j - \sqrt{n_i n_j}\,\bigr] \int \chi_{ij}(r_1) \,
\chi_{ji}(r_2) \, (r_1 - r_2)^2 \,dr_1 \,dr_2
\\
&= - \sum_{i=0}^\infty \bigl[ (n_i^2 - n_i) L_i
+ 2(n_i n_{i+1} - \sqrt{n_i n_{i+1}}\,) M_i \bigr]
\\
&= \frac{2\om t}{(1 + t)^{2}} \biggl[ \frac{1 - t^2}{1 + t^2}
- \frac{1 + t^2}{(1 + t)^2} \biggr],
\end{align*}
or equivalently,
$$
E_\ee^\HF(\om,\mu) - E_\ee^\Mu(\om,\mu)
= \frac{\om - \mu}{\om + \mu}\, \sqrt{\om\mu}
- \frac{\om^2 - \mu^2}{4\om} \,.
$$
At $t = 0$ there is no difference between these two values of the
energy. It is worth noting that there is another point of coincidence,
namely $t \sim 0.54$ or $\dl/k \sim 0.99$. Below this value the
difference is positive, and above it is negative. At $t = 1$, we find
$$
E_\ee^\HF(\om,0) - E_\ee^\Mu(\om,0) = - \frac{\om}{4} \,.
$$

Since $\rho_1^\HF - \rho_1 = 2 \sum_i (n_i - n_i^2)\chi_i \neq 0$ for
$t > 0$, this functional does not satisfy the sum rule, except when
the Hartree--Fock functional is evaluated on a Hartree--Fock state.

\subsection{The Goedecker--Umrigar functional} 
\label{ssc:GU-fnal}

The Goedecker--Umrigar functional~\cite{GoedeckerU} introduces a small
variation of M\"uller's, attempting to exclude ``orbital
self-interaction''. For our closed-shell situation, it is given by:
\begin{align*}
\rho_2^\GU(r_1,r_2) - \rho_2^\Mu(r_1,r_2)
= \sum_i (n_i - n_i^2) \,\chi_i(r_1) \,\chi_i(r_2).
\end{align*}
This relation implies that for $t>0$ this functional violates the sum
rule: $\rho_1 \neq 2\int \rho^\GU_2 \,dr_2$. The interelectronic part
of the energy difference calculation is given by
$$
\sum_r (n_i - n_i^2) L_i 
= \sum_i \bigl[ (1 - t^2) t^{2i} - (1 - t^2)^2 t^{4i} \bigr] L_i \,.
$$
Hence, the mean value of the interelectronic repulsion predicted by
this functional is
$$
E_\ee^\GU(t) - E_\ee^\Mu(t) = \frac{2\om t}{(1 + t)^2} \biggl[
\frac{1 + t^4}{1 - t^4} - \biggl(\frac{1 + t^2}{1 - t^2}\biggr)^{\!2}
\biggr].
$$
The interelectronic energy calculated by means of the
Goedecker--Umrigar functional is higher than the exact value. At
$t = 1$, the difference diverges. This is unsurprising, given that
when the coupling is large enough the self-interacting part is almost
half of the total interelectronic energy; for instance,
$E_\ee^\GU(0.8) / E_\ee(0.8) \sim 0.44$.

\vspace{6pt}

\begin{table}[ht] 
\centering      
\begin{tabular}{l c c c}
\hline\noalign{\smallskip}
2-RDM  & Antisymmetry & Hermiticity & Sum Rule \\ [0.5ex] 
\noalign{\smallskip}\hline\noalign{\smallskip}         
Exact         & \yea & \yea & \yea \\  
M\"uller      & \nay & \yea & \yea \\
Hartree-Fock   & \yea & \yea & \nay \\  
GU            & \nay & \yea & \nay \\ 
BBC           & \nay & \yea & \yea \\ 
CHF           & \nay & \yea & \yea \\   
CGA           & \nay & \yea & \yea \\ [1ex]   
\noalign{\smallskip}\hline   
\end{tabular} 
\caption{Properties fulfilled by the exact two-body functional for
two-electron atoms and several two-body reduced density approximations
\cite{Helbig06}.}
\label{table:props} 
\end{table}

\subsection{Buijse--Baerends corrected functionals} 
\label{ssc:BB-fnals}

A few years after the original Buijse and Baerends'
paper~\cite{BuijseB02}, some corrections were introduced, to
distinguish between strongly occupied natural orbitals (whose
occupation numbers are close to~$1$) and weakly occupied ones
(occupation numbers near~$0$) \cite{GritsenkoPB05}. The harmonium
ground state possesses only one strongly occupied orbital, namely
$\chi_0$, whose occupation number is $n_0 = 1 - t^2$. However, this
distinction is lost at high values of the coupling parameter. The
first corrected functional (BBC1) is given by
$\rho_2^\BBcuno = \rho_2^\Mu + \rho_2^\Cuno$, where
\begin{align*}
\rho_2^\Cuno(r_1,r_2) = 2 \sum_{\substack{i \neq j \\ i,j > 0}}
\sqrt{n_i n_j}\, \chi_{ij}(r_1) \, \chi_{ji}(r_2).
\end{align*}
The second correction (BBC2) modifies BBC1 by adding further terms
of the form 
$(\sqrt{n_i n_j} - n_i n_j) \,\chi_{ij}(r_1) \,\chi_{ji}(r_2)$ for 
distinct strongly coupled orbitals. For the harmonium ground state,
we may ignore it here; thus we write $\rho_2^\BBc$ for~$\rho_2^\BBcuno$.
Since both corrections involve only off-diagonal terms ($r \neq s$),
these corrected functionals still fulfil the sum rule. 

\begin{figure}[ht] 
\centering
\includegraphics[width=12cm]{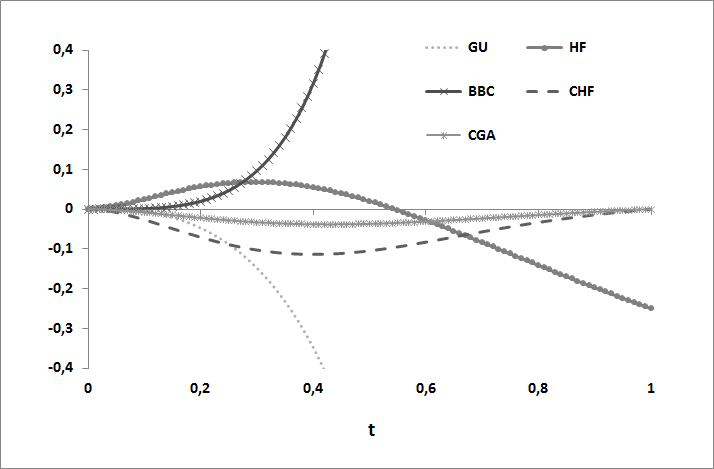} 
\caption{The error of the energy value calculated for different
functionals. The error is defined as the dimensionless
$[E_\ee^\mathrm{functional}(t) - E_\ee(t)]/\om$, evaluated on the
exact one-body density matrix for the harmonium ground state. The
M\"uller functional does not appear here since its energy value is
exact.}
\label{graf:energy-diff} 
\end{figure}

The functional difference now reads
$\rho_2^\BBc - \rho_2 = \rho_2^\Cuno$ and the interelectronic energy
difference yields
\begin{align*}
E_\ee^\BBc - E_\ee^\Mu
&= \frac{\mu^2 - \om^2}{4} \int \rho_2^\Cuno \, r^2_{12} \,dr_1\,dr_2
= 4 \sum_{i>0} \sqrt{n_i n_{i+1}}\, M_i
\\
&= \frac{\om^2 - \mu^2}{\sqrt{\om\mu}}\, 
\sum_{i>0} \sqrt{n_i n_{i+1}}\, (i + 1)
= \frac{8\om t^4}{(1 + t)^2}\, \frac{1 + t^2}{1 - t^2}
\biggl[ \frac{1}{1 - t^2} + 1 \biggr].
\end{align*}
As in the Goedecker--Umrigar functional case, at $t = 1$ the
difference has a divergence. Over almost the whole range of~$t$, there
is a large error in the energy (see Figure~\ref{graf:energy-diff}).
Thus, applied to harmonium, these functionals do not reproduce the
success found for the homogeneous electron gas~\cite{LathiotakisHG07}.

\subsection{CHF and CGA functionals} 
\label{ssc:CHF-CGA-fnals}

\begin{figure}[t] 
\centering
\includegraphics[width=12cm]{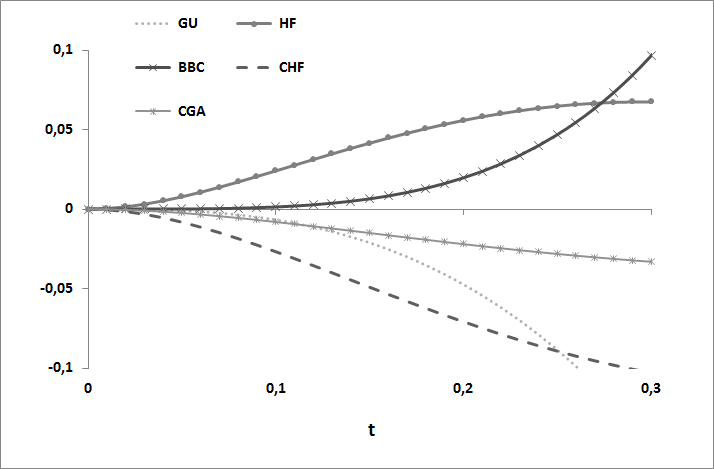} 
\caption{A zoom of the energy value errors from
Figure~\ref{graf:energy-diff}.}
\label{graf:diff-zoom} 
\end{figure}

Corrected Hartree--Fock (CHF) and Cs\'anyi--Goedecker--Arias (CGA)
functionals introduced in~\cite{CsanyiGA02} are improvements of the
Hartree--Fock functional. They were designed as tensor products to get
better predictions for the correlation energy in homogeneous electron
gases at high densities. For a closed shell system, they read
\begin{align*}
d_2^\CHF(u_1,u_2) &= \frac{1}{2}\, d_1(u_1)\,d_1(u_2) - \sum_{i,j}
\Bigl( n_i n_j + \sqrt{n_i(1 - n_i) n_j(1 - n_j)}\, \Bigr)
\chi_{ij}(u_1) \,\chi_{ji}(u_2),
\\
d_2^\CGA(u_1,u_2)
&= \frac{1}{2}\, d_1(u_1)\,d_1(u_2) - \frac{1}{2} \sum_{i,j}
\Bigl( n_i n_j + \sqrt{n_i(2 - n_i) n_j(2 - n_j)}\, \Bigr)
\chi_{ij}(u_1) \,\chi_{ji}(u_2).
\end{align*}
First, note that both functionals satisfy the sum rule:
$d_1^\CHF = d_1^\CGA = d_1$. As regards the interelectronic energy,
we find that
\begin{align*}
E_\ee^\CHF - E_\ee^\Mu
&= 2 \sum_i \sqrt{n_i\,n_{i+1}} \bigl( 1 - \sqrt{n_i\,n_{i+1}}
- \sqrt{(1 - n_i)(1 - n_{i+1})}\, \bigr) M_i,
\\
E_\ee^\CGA - E_\ee^\Mu
&= \phantom{2} \sum_i \sqrt{n_i\,n_{i+1}} \bigl(
2 - \sqrt{n_i\,n_{i+1}} - \sqrt{(2 - n_i)(2 - n_{i+1})}\, \bigr) M_i.
\end{align*}
As can be seen in Figures \ref{graf:energy-diff}
and~\ref{graf:diff-zoom}, both functionals show a remarkably good
description of the energy. At $t= 0$ and $t = 1$ the energy is exact.
For the CHF functional, the worst performance occurs around
$t \sim 0.4$ or $\dl/k \sim 0.96$, whose error is
$(E_\ee^\CHF - E_\ee^\Mu)/\om \sim 0.11$; the CGA functional is worst
at $t \sim 0.43$ or $\dl/k \sim 0.97$, with an error of
$(E_\ee^\CHF - E_\ee^\Mu)/\om \sim 0.03$. The estimates of the energy
are lower than the correct one; thus they are both overbinding for
harmonium.

\section{Conclusion}
\label{sec:conclusion}

We have used harmonium as a laboratory to study the performance of
some of the one-body density functionals proposed to compute the
interelectronic repulsion energy in the framework of DMFT. We have
confirmed the exact value of the energy given by the M\"uller
functional when evaluated on the exact ground state. The functionals
which exclude self-interacting terms or distinguish between strongly
and weakly occupied orbitals display good approximation for the energy
at small values of the coupling parameter but very poor values beyond
$t \sim 0.3$. 

The CHF approximation yields a good description of the interelectronic
repulsion, even at high values of the parameter $t$. The performance
of the CGA approximation is remarkably good, taking into account that
it was built explicitly for the electron gas case. The reader should
keep in mind the violation of some physical constraint or other by
each one of the examined functionals \cite{LathiotakisGH10}.

\subsection*{Acknowledgments}

We are most grateful to Jos\'e M. Gracia-Bond\'ia for a careful
reading of the manuscript. CLBR has been supported by a Banco
Santander scholarship. JCV thanks the Departamento de F\'isica
Te\'orica of the Universidad de Zaragoza for warm hospitality, and
acknowledges support from the Direcci\'on General de Investigaci\'on e
Innovaci\'on of the regional government of Arag\'on, and from the
Vicerrector\'ia de Investigaci\'on of the University of Costa Rica.


\end{document}